\documentclass[polymers,article,accept,moreauthors,pdftex,10pt,a4paper]{mdpi}

\firstpage{1}
\articlenumber{x}
\doinum{10.3390/------}
\pubvolume{x}
\pubyear{year}
\copyrightyear{year}
\externaleditor{Academic Editor: name}
\history{Received: date; Accepted: date; Published: date}

\theoremstyle{mdpi}
\newcounter{thm}
\newcounter{ex}
\newcounter{re}
\usepackage{amssymb}
\usepackage{amsmath}
\usepackage{bm}
\usepackage{wasysym}
\usepackage{hyperref} 
\usepackage{soul}

\newcommand{\lla}{\left\langle}
\newcommand{\rra}{\right\rangle}

\graphicspath{{../figures/}}

\newcommand{\RW}[1]{\textcolor{black}{#1}}

\Title{Tethered Semiflexible Polymer under Large Amplitude Oscillatory Shear}

\Author{ {Antonio Lamura} $^{1,}$* and Roland G. Winkler $^{2,}$*\href{https://orcid.org/0000-0002-7513-0796}{\orcidicon}}

\AuthorNames{Antonio Lamura   and Roland G. Winkler  }

\address{%
$^{1}$ \quad Istituto Applicazioni Calcolo, CNR, Via Amendola 122/D, 70126 Bari, Italy \\
$^{2}$ \quad Theoretical Soft Matter and Biophysics, Institute for
Advanced Simulation,
Forschungszentrum J\"ulich, D-52425 J\"ulich, Germany}

\corres{Correspondence: antonio.lamura@cnr.it (A.L.); r.winkler@fz-juelich.de (R.G.W.)}

\abstract{The~properties of a semiflexible polymer with fixed ends exposed to oscillatory shear flow are investigated by simulations. The~two-dimensionally confined polymer is modeled as a linear bead-spring chain, and the interaction with the fluid is described by the Brownian multiparticle collision dynamics approach.
For small shear rates, the tethering of the ends leads to a more-or-less linear oscillatory response. However, at high shear rates, we found a strongly nonlinear reaction, with a polymer (partially) wrapped around the fixation points. This leads to an overall shrinkage of the polymer. Dynamically, the location probability of the polymer center-of-mass position is largest on a spatial curve resembling a lima\c{c}on, although with an inhomogeneous distribution. We found shear-induced modifications of the normal-mode correlation functions, with a frequency doubling at high shear rates. Interestingly, an even-odd asymmetry for the Cartesian components of the correlation functions appears, with rather similar spectra for odd $x$- and even $y$-modes and vice versa. Overall, our simulations yielded an intriguing nonlinear behavior of tethered semiflexible polymers under oscillatory shear flow.}

\keyword{mesoscale simulations; nonequilibrium simulations; LAOS; polymer dynamics}

\begin{document}

\section{Introduction}
The~structural and rheological properties of polymers are strongly affected by external fields, such as shear or extensional flows. As such, this is well established, and the external fields provide means to control the behavior of polymer solutions, melts, and networks~\cite{bird:87.1,rubi:03,lars:15,shaw:18,schr:18}. Similarly, it has been shown that the macroscopic rheological behavior of a polymer solution, e.g., shear-rate dependent viscosities, normal-stress differences, and shear thinning, are tightly linked to the microscopic polymer conformational and dynamical properties~\cite{doi:86,bird:87,oett:96,prak:01,huan:10,wink:10,schr:18}. Hence, insight into the behavior of individual polymers is fundamental in the strive to unravel the macroscopic nonequilibrium polymer properties.

Direct observation of the nonequilibrium properties of single molecules, also termed ``molecular rheology''~\cite{schr:18}, in experiments~\cite{smit:99,schr:05,teix:05,schr:05.1,doyl:00,lado:00,gera:06,zhou:16} and simulations~\cite{liu:89,delg:06,hur:00,jose:08,he:09,knud:96,lyul:99,pete:99,jend:02,hsie:04,liu:04,pami:05,schr:05,send:08,pier:95,
aust:99,grat:05,zhan:09,ryde:06,ripo:06,koba:10,huan:10,lamu:12,wink:14.1,cann:08,chel:10,chel:12}
has provided valuable visual illustrations of polymer conformations and has helped to characterize their nonequilibrium properties in terms of their deformation, orientation, relaxation dynamics, and rheology, both for free and tethered molecules in shear and extensional flow. These studies, however, typically probe the linear viscoelastic properties, which are usually insufficient to fully characterize the nonlinear aspects of polymers under~flow.

In order to probe nonlinear properties of complex fluids, the large-amplitude oscillatory shear (LAOS) method was developed~\cite{bogi:66,phil:66,giac:93,hyun:11,zhou:16}. Here, in contrast to small-amplitude oscillatory shear, the stress response is typically no longer sinusoidal, but of rather complex shape.
Such strong, time-dependent flows will affect the conformational properties of individual polymers to a yet unresolved extent. Studies on the dynamics of single polymers under large-amplitude oscillatory extensional flow yield qualitatively different stretching flow-rate curves (Lissajous curves) as a function of the extension rate and the oscillation frequency~\cite{zhou:16}, and illustrate the complex interplay between time-dependent flows and polymer conformations. Here, further studies are desirable to resolve the time-dependent conformational properties of single polymers under large-amplitude oscillatory flows, aspects which, so far, have not been addressed by simulations.

In this article, we perform non-hydrodynamic, Brownian-type simulations of individual polymers exposed to large-amplitude oscillatory shear. The~ends of the polymer are fixed and the polymer dynamics is constraint to the $xy$-plane of a Cartesian reference frame. We consider stiffer polymers only, with the persistence lengths $L_p/L = 1/2 $ and $2$, respectively ($L_p$ is the persistence and $L$ the polymer contour length). In any case, the polymers exhibit cyclic conformational modulations, specifically at higher Weissenberg numbers ($Wi$), \RW{which is the product of the applied shear rate and the longest polymer relaxation time.} The~tethering of the ends leads to a more-or-less linear oscillatory response at small Weissenberg numbers, where a polymer moves back and forth like grass swaying in the wind. With increasing $Wi$, polymers (partially) wrap around the fixation points and shrink. In general, the probability of the center-of-mass position is largest on a lima\c{c}on-type curve, as a consequence of the periodic excitation, however, with a non-uniform probability. Interestingly, the center-of-mass autocorrelation function normal to the line connecting the tethering points exhibits frequency doubling with respect to the imposed shear frequency, as a consequence of the non-crossability of the polymer and the tethering points. This reflects the symmetry breaking of the polymer dynamics during an oscillation cycle. Our findings illustrated the complex nonlinear interplay of polymer internal degrees of freedom and external periodic oscillations.

The~rest of the paper is organized as follows. In Section \ref{sec:model}, the polymer model and the coupling to the shear flow are described. Section \ref{sec:results} presents results for conformational and dynamical properties. Our findings are summarized in Section \ref{sec:conclusions}.

\section{Model and Method} \label{sec:model}

\begin{figure}[H]
\centering
\includegraphics*[width=0.5\columnwidth,angle=0]{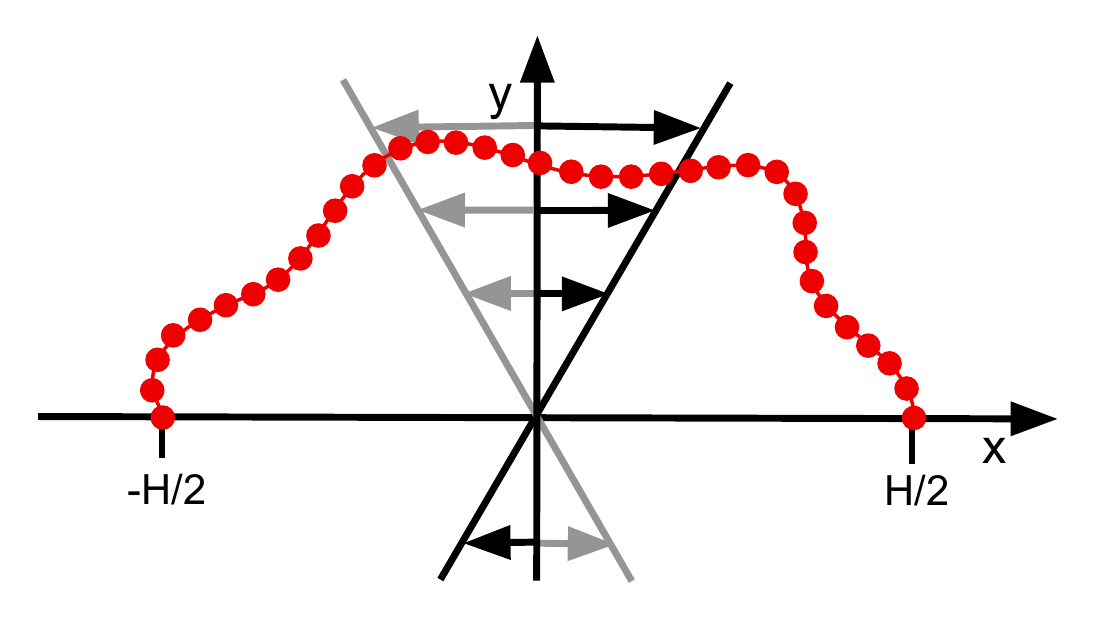}
\caption{Sketch of the tethered bead-spring polymer exposed to oscillatory linear shear flow.
\label{fig:sketch}
}
\end{figure}

The~two-dimensional linear polymer chain is composed of $N$ beads of mass $M$, with its ends, $\bm r_1$, $\bm r_N$, tethered at $\bm r_1 = (-H/2,0)^T$
and $\bm r_N =(H/2,0)^T$, respectively \RW{(cf. Figure~\ref{fig:sketch})}.  The contour length $L=(N-1)r_0$, where $r_0$ is the bond length, is fixed and $L>H$.
The~interactions between the beads are defined in terms of the potential
$U=U_{bond}+U_{bend}+U_{ex}$, comprising bond, bending, and excluded-volume interactions.
The~bonds between consecutive beads are described by the harmonic~potential:
\begin{equation}\label{bond}
U_{bond}=\frac{\kappa_h}{2} \sum_{i=1}^{N-1}
(|{\bm r}_{i+1}-{\bm r}_{i}|-r_0)^2 ,
\end{equation}
where ${\bm r}_i$ is the position of bead $i$
($i=1,\ldots,N$) and $\kappa_h$ is the elastic constant.
Bending restrictions are captured by the potential:
\begin{equation}
U_{bend}=\kappa \sum_{i=1}^{N-2} (1-\cos \varphi_{i}) ,
\label{bend}
\end{equation}
with $\kappa$ the bending rigidity and $\varphi_{i}$ the angle between two
consecutive bond vectors. In the limit $\kappa \to \infty$, the persistence
length is given by $L_p=2 \kappa r_0/ k_B T$, with $T$ the temperature and $k_B$ Boltzmann's constant.
Bead overlapping and bond crossings are prevented by the shifted and
truncated Lennard--Jones~potential:
\begin{equation}
U_{ex} =
4 \epsilon \Big [ \Big(\frac{\sigma}{r}\Big)^{12}
-\Big(\frac{\sigma}{r}\Big)^{6} +\frac{1}{4}\Big] \Theta(2^{1/6}\sigma -r) ,
\label{rep_pot}
\end{equation}
where $r$ is the distance between two non-bonded beads, and
$\Theta(r)$ is the Heaviside function; $\Theta(r)=0$ for $r<0$ and
$\Theta(r)=1$ for $r > 0$. The~dynamics of the beads is
described by Newton's equations of motion, which are integrated by
the velocity-Verlet algorithm with time step $\Delta t_p$
\cite{swop:82,alle:87}.

The~polymer is coupled to a Brownian heat bath implemented
via the Brownian (or random) multiparticle collision dynamics (B-MPC) approach
\cite{ripo:07,gomp:09,kiku:02}. Hence, no hydrodynamic effects
are considered in the present work. MPC consist of streaming and collision steps, where collisions occur in regular time intervals of length $\Delta t$~\cite{kapr:08,gomp:09}. During streaming, the dynamics of the beads is described by Newton's equations of motion. In the collision step, the velocities of the beads change in a stochastic manner.
In B-MPC, the Brownian interaction of a bead with the surrounding fluid is implemented by a stochastic collision with a phantom particle, taking its
momentum from the Maxwell--Boltzmann distribution of variance $M k_B T$ and a mean, which is zero in absence of shear. In the presence of oscillatory shear in the $xy$-plan the mean momentum is $M \bm v_i^s(t)$, with the shear velocity:
\begin{align}
\bm v_i^s(t) =  \left(\dot{\gamma} y_{i} \cos(\omega t ),0 \right)^T
\end{align}
at the time $t$; $\dot \gamma$ is the shear rate and $\omega$ the frequency \RW{(cf. Figure~\ref{fig:sketch})}. The~collision is implemented via the stochastic rotation dynamics variant of MPC~\cite{ihle:01,lamu:01,gomp:09}. Here, the
relative velocity of a bead, with~respect to the mean of the velocities of the bead and related phantom
particle, is randomly rotated by angles $\pm \alpha$.

We choose the following parameters for the simulations: $\alpha=130^{\circ}$, $M=5$,
$\Delta t=0.1 t_u$, with the time unit $t_u=\sqrt{M r_0^2/( 5 k_B
T)}$, $\kappa_h r_0^2/(k_B T)=4~\times~10^3$, $\epsilon /
(k_B T)=1$, $\sigma=r_0$, $N=101$, hence, the~polymer length is $L=100
r_0$, and $\Delta t_p=10^{-2} \Delta t$. The~value of
$\kappa_h$ guarantees that the polymer length is constant within
$1\%$ for all considered systems. Two bending rigidities are considered, corresponding to the persistence lengths $L_p/L = 0.5$ and $2$.

Simulations of free polymers yield the longest relaxations times $\tau_r/t_u = 1.9~\times~10^6$ and $\tau_r/t_u = 3.9~\times~10^6$ for the two stiffness values, determined from the end-to-end vector correlation function~\mbox{\cite{wink:10,lamu:12}}.
The~strength of the shear flow is characterized by the Weissenberg number $Wi=\tau_r \dot{\gamma}$, for which the values $Wi=10, \,  25, \, 50,$ and $100$ are considered. The~frequency $\omega$ is related to the Deborah number $De=\tau_r \omega$, where we set $De=10$.

The~polymer, with $H = L/5$, is initialized with beads along a semi-elliptical contour
with the major axis along the $y$-direction and minor axis along the $x$-axis. The~polymers are equilibrated up to $5~\times~10^6 t_u > \tau_r$, and
data are collected up to the longest simulated time $t_L=10^{8} t_u$.\vspace{-6pt}

\section{Results} \label{sec:results}

\begin{figure}[H]
\centering
\includegraphics*[width=0.75\textwidth,angle=0]{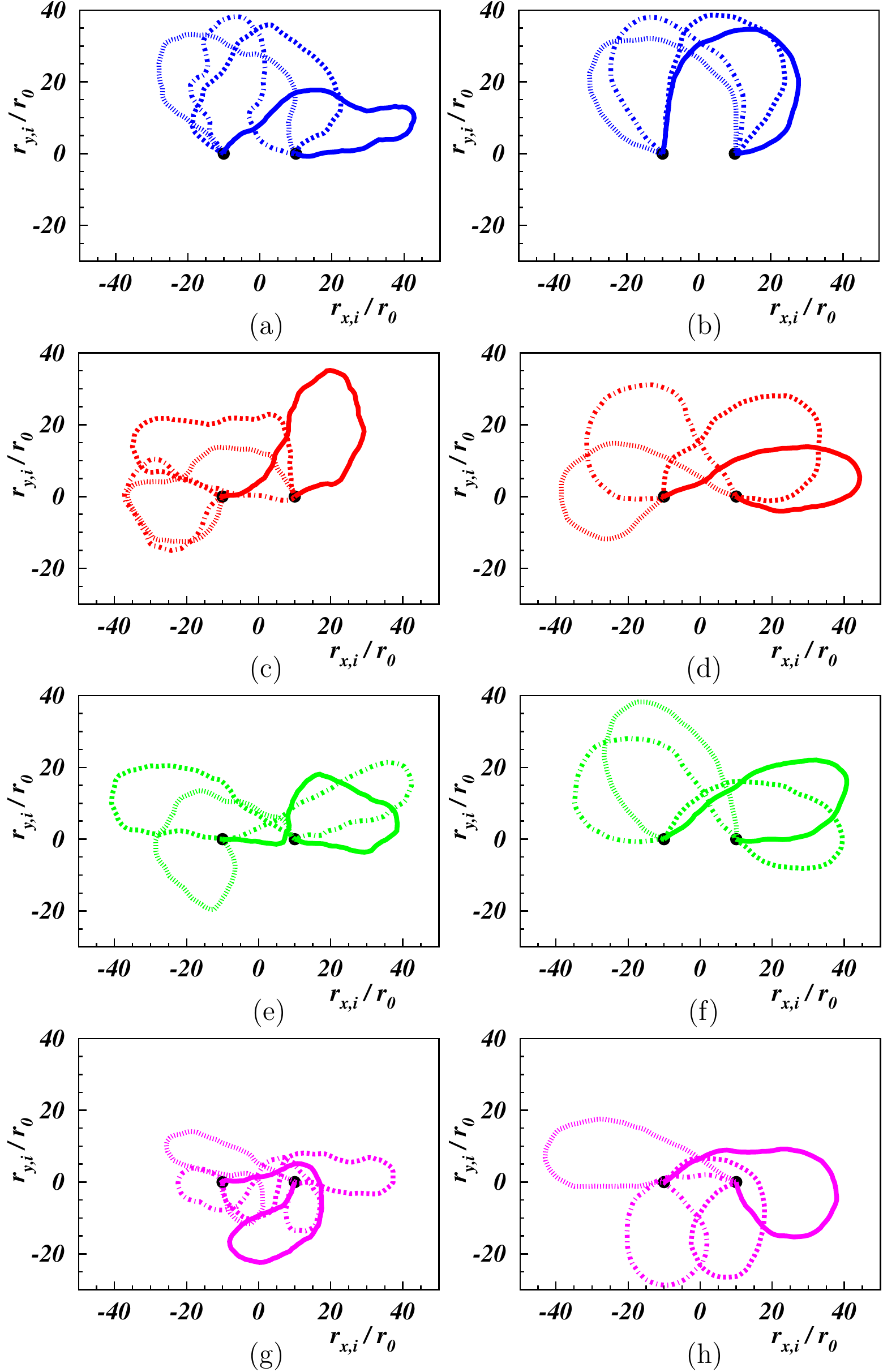}
\caption{ {Conformations} of the polymer at times
$t= 0$ (solid line), $0.5\pi/\omega$ (dashed line),
$\pi/\omega$ (dotted line),
$1.5\pi/\omega$ (dash-dotted line) after equilibration
for $L_p/L=0.5$ (\textbf{left}), $2$ (\textbf{right}), and
$Wi$ = 10, 25, 50, 100~(\textbf{from top to bottom}).
Black dots denote the position of fixed beads.
\label{fig:snapshots}
}

\end{figure}

The~oscillatory flow dragged the beads along and, at least at small shear rates, the polymer moved back and forth like grass swaying in the wind. This is illustrated in Figure~\ref{fig:snapshots}. The~stiffer polymer ($L_p/L=2$) closely maintained its shape at low shear rates (Figure~\ref{fig:snapshots}b). A flexible polymer was deformed more easily, but an in-phase oscillation was still present (Figure~\ref{fig:snapshots}a). Larger Weissenberg numbers led to stronger conformational changes and larger bead displacements (Figure~\ref{fig:snapshots}c--h).

\subsection{Center-of-Mass Properties}

\begin{figure}[H]
\centering
\includegraphics*[width=0.6\columnwidth,angle=0]{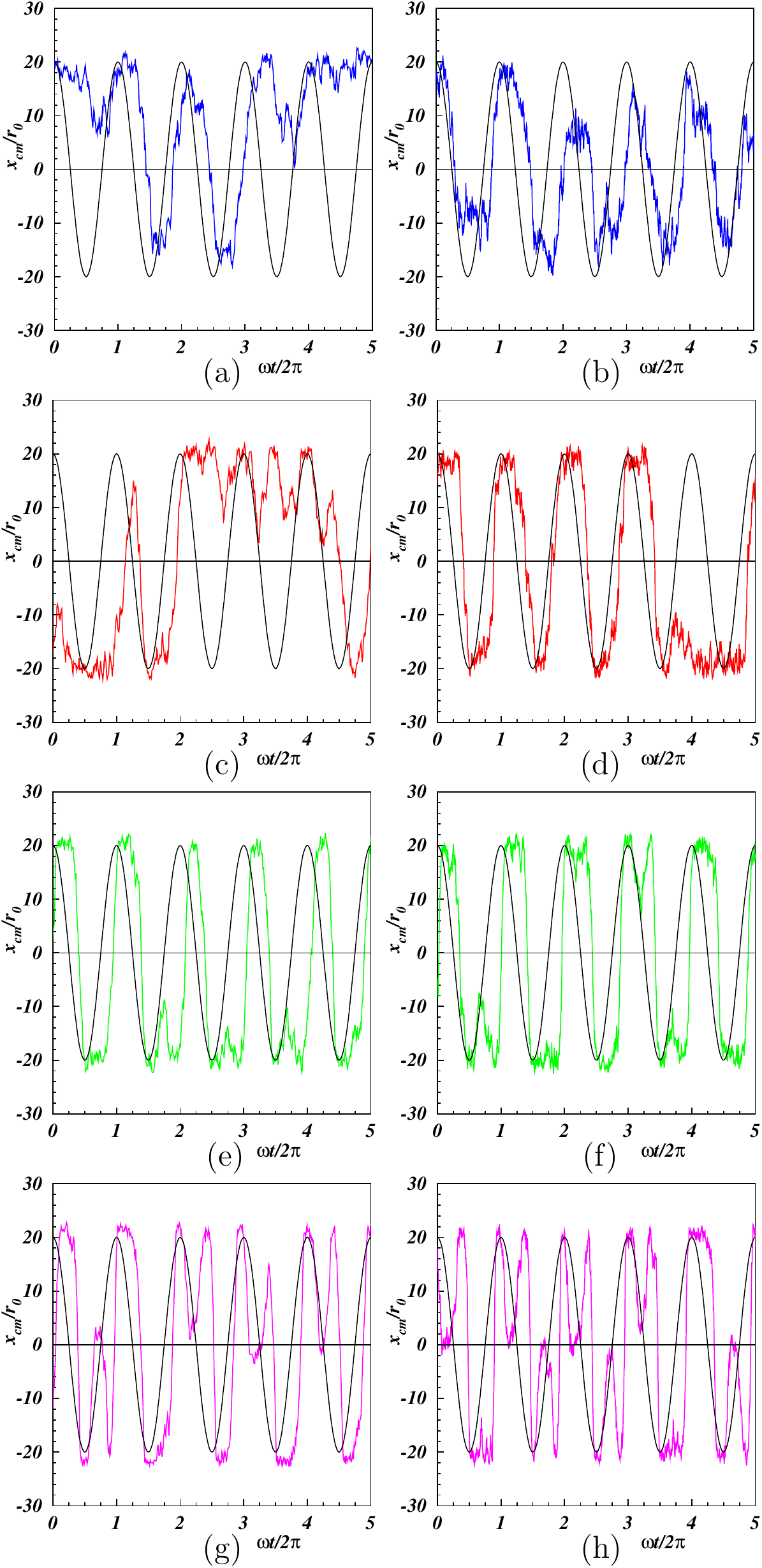}
\caption{ {Time-dependence} of the center-of-mass coordinate along the $x$-axis for $L_p/L=0.5$ (\textbf{left}) and $2$ (\textbf{right}), and $Wi$ = 10, 25, 50, 100~(\textbf{from top to bottom}). The~black lines correspond to the externally applied shear flow
with arbitrary amplitude.
\label{fig:time_dependence_xcm}
}
%
\end{figure}

Figure~\ref{fig:time_dependence_xcm} shows the time-dependence of the $x$-coordinate of the center-of-mass position for various Weissenberg numbers. For smaller $Wi \lesssim 25$, entropic effects were strong, and $x_{cm}$ was only partially following the external flow. Perturbations were stronger when the polymer got trapped by the
fixation points and some time was needed to disentangle it (cf.
Supplementary Movies for $L_p/L=0.5, 2$ and $Wi=25$).
There were in-phase periods with a small phase shift, which were interrupted by time intervals with a center-of-mass motion decoupled from the flow. A stronger flow ($Wi \approx 50$) enhanced the in-phase periodic motion, but the center-of-mass dynamics was phase shifted and seemed to be no longer harmonic. The~modulations of the approximately periodic peaks became more pronounced for $Wi=100$, and an original single peak split into two peaks, with the minimum of the second peak close to zero. Hence, $x_{cm}$ exhibited an approximate doubling of the frequency, an aspect which is more closely discussed in Section~\ref{sec:dynamics} in the context of normal modes. Overall, we found a highly nonlinear response of the polymer to the external excitation. This modified the polymer conformational and dynamical properties (cf. Supplementary Movies for $L_p/L=0.5, 2$ and $Wi=100$).

\begin{figure}[H]
\centering
\includegraphics*[width=0.8\columnwidth,angle=0]{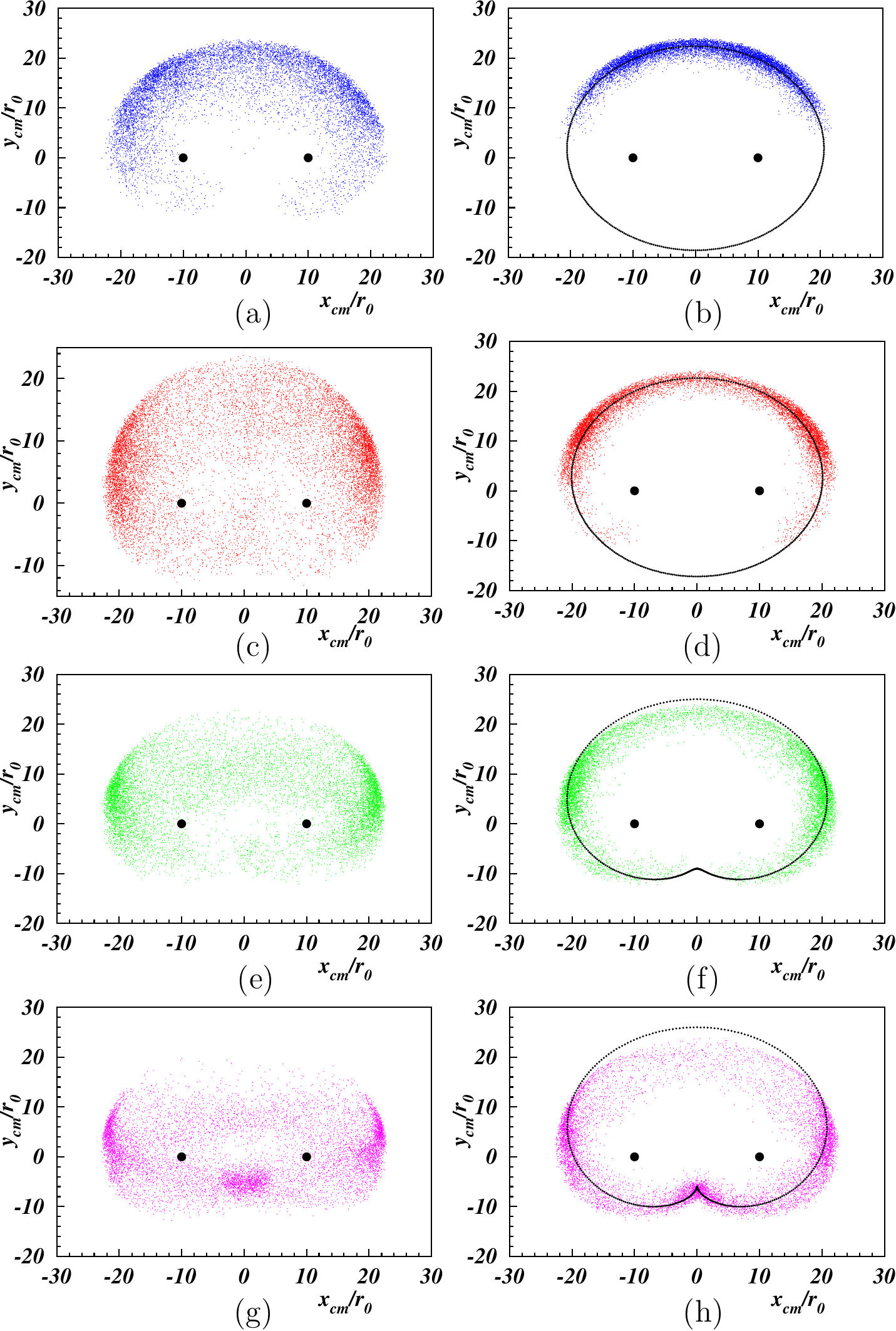}
\caption{ {Probability} distribution of the center-of-mass position of the tethered polymers
for $L_p/L=0.5$ (\textbf{left}) and $2$ (\textbf{right}), and
$Wi=10, 25, 50, 100$ (\textbf{from top to bottom}).
Black dots indicate the position of fixed ends. \RW{The~lines for $L_p/L=2$ (right) are lima\c{\textbf{c}}ons according to Equations~\eqref{eq:lim_x} and \eqref{eq:lim_y} for the parameters: (\textbf{b}) $a/r_0=20.48$, $b/r_0=1.92$, $y_0=0$, (\textbf{d}) $a/r_0=19.91$, $b/r_0=2.72$, $y_0=0$, (\textbf{f}) $a/r_0=17$, $b/r_0=14$, $y_0/r_0=-6$, (\textbf{h}) $a/r_0=16$, $b/r_0=16$, $y_0/r_0=-6$.}
\label{fig:prob_cm_pos}
}
%
\end{figure}

Figure~\ref{fig:prob_cm_pos} depicts the probability distribution of the center-of-mass position for various Weissenberg numbers. The~probability was high for positive $y_{cm}$-values, specifically at lower $Wi$. This was related to the chosen initial conditions, with a polymer always in the half-plane $y>0$. Evidently, the~polymers were too stiff to restore isotropy normal to the shear-flow direction. The~anisotropy was maintained at higher $Wi$. However, the most probable center-of-mass position shifted gradually to larger $|x_{cm}|$. At large Weissenberg numbers, the probability increased in the vicinity of $x_{cm}=0$. As~illustrated in Figure~\ref{fig:snapshots}g,h, the polymers were wrapped around the tethering points by the flow. Overall, the~probability of the center-of-mass position was highest on a lima\c{c}on-type curve, however, with a non-uniform distribution. A lima\c{c}on is defined as a curve formed by the path of a point fixed to a circle, when that circle rolls around the outside of a circle of equal radius. In our case, the shear flow (partially) rotated (oscillated) the semiflexible polymer of more-or-less circular shape, which looked like rolling, and the center-of-mass followed a lima\c{c}on.

\RW{The~polar coordinate representation of a lima\c{c}on is $r=a+b\cos \theta$, and the parameter representation for our reference frame is~\cite{bron:73}:
\begin{align} \label{eq:lim_x}
x = &~r \cos \theta  = \left[a+b \sin \theta  \right] \cos \theta , \\ \label{eq:lim_y}
y = &~y_0+r \sin \theta= y_0 + \left[a+b \sin \theta  \right] \sin \theta ,
\end{align}
with the off-set $y_0$. The~fits of Equations~\eqref{eq:lim_x} and \eqref{eq:lim_y} to our simulation results for $L_p/L=2$ are displayed in Figure~\ref{fig:prob_cm_pos}. For $Wi=10$, the lima\c{c}on was hardly distinguishable from an ellipse. With increasing Weissenberg number, we approached a lima\c{c}on, and for $Wi=100$, the lima\c{c}on turned into a cardioid, where $a=b$. The~lima\c{c}on curves for $Wi=50$, and especially $Wi=100$, did not fully agree with the simulation data for $y_{cm}/r_0 \gtrsim 10$. The~reason was the conformational freedom of the polymer and shear-induced shape changes, implying changes in the radii of the rolling circles underlaying the mathematical lima\c{c}on construction. These changes were most pronounced while the polymer explored regions of high shear rate.
}

\begin{figure}[H]
\centering
\includegraphics*[width=0.75\columnwidth,angle=0]{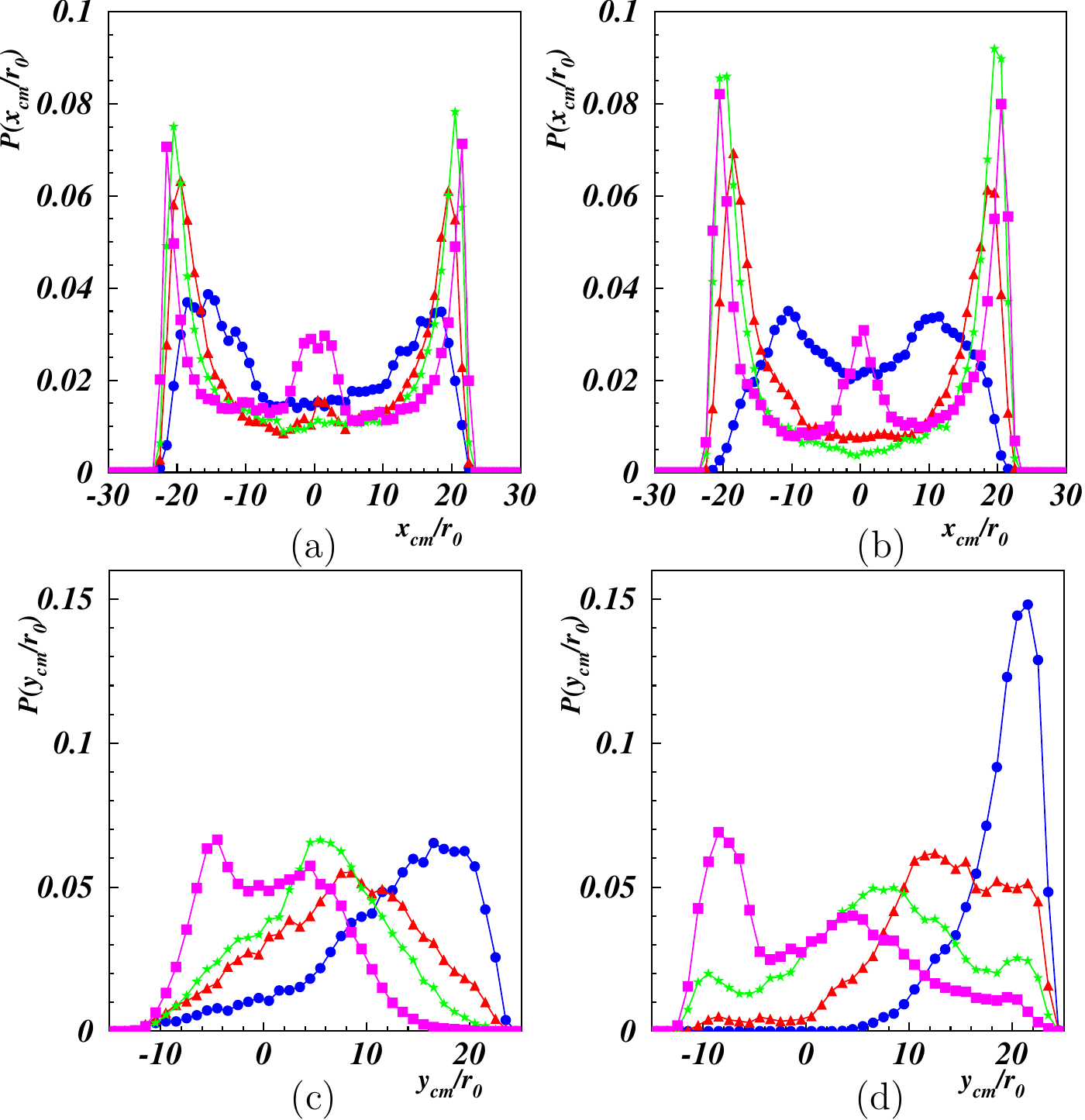}
\caption{ {Probability} distribution functions
of the $x$ (\textbf{top}) and $y$ (\textbf{bottom}) coordinates of the polymer center of mass
for $L_p/L=0.5$ (\textbf{left}), $2$ (\textbf{right}) and $Wi=10~(\bullet), 25~(\blacktriangle),
50~(\star), 100~(\blacksquare)$.
\label{fig:prob_cm_comp}
}
%
\end{figure}

\begin{figure}[H]
\centering
\includegraphics*[width=0.5\columnwidth,angle=0]{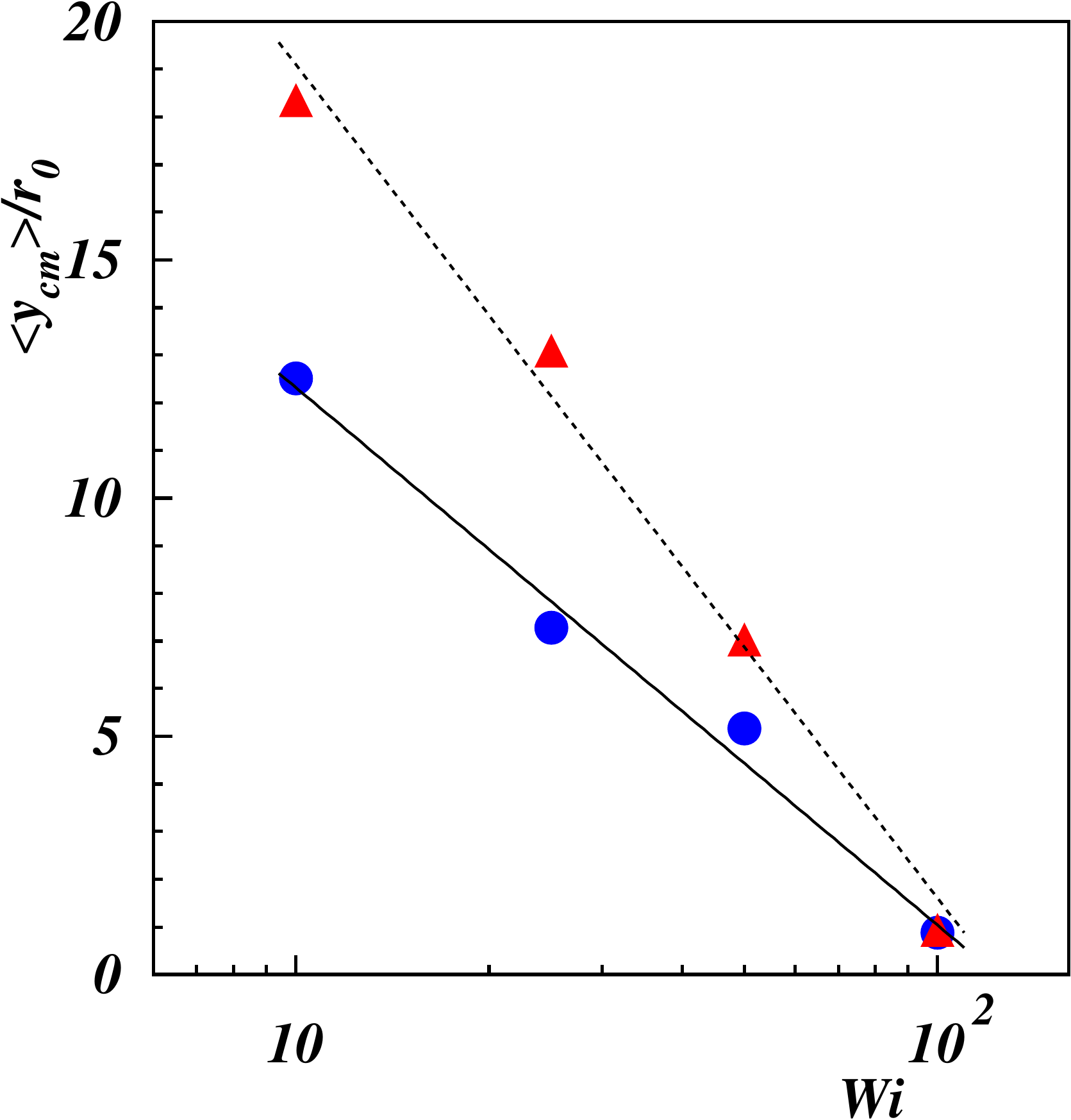}
\caption{Average position along the $y$-direction
of the polymer center of mass as a function of the Weissenberg number
for $L_p/L=0.5$ ($\bullet$), $2$ ($\blacktriangle$). Lines are the fits
$<y_{cm}>/r_0 \sim \beta \ln Wi$
with $\beta=-4.9 \pm 0.4$ (full line)
$\beta=-7.6 \pm 0.6$ (dashed line).
\label{fig:ave_y_cm}
}
\end{figure}

The distribution functions for the center-of-mass Cartesian coordinates of the polymers are displayed in Figure~\ref{fig:prob_cm_comp}. The distribution function for $P(x_{cm})$ (Figure~\ref{fig:prob_cm_comp}a,b{)} is symmetric with respect to the center between the tethered ends. At small shear, distinct ``off-center'' peaks were present as a consequence of the projection onto the $x$-axis. With the increasing Weissenberg number, the~peaks became more pronounced and shift closer to the end positions. This is also reflected in the variance $\lla x_{cm}^2 \rra$, which increased with $Wi$. At high Weissenberg numbers, a peak appeared in the center, reflecting the polymer ``wrapping'' around the fixed ends. The asymmetry in the initial condition of $y_{cm}$ is also reflected in Figure~\ref{fig:prob_cm_comp}c,d, with a pronounced peak at positive $y_{cm}$. The probability of smaller $y_{cm}$ increased with increasing $Wi$, and for $Wi \gtrsim 100$, a peak appeared at $y_{cm} <0$, consistent with the high probability in the vicinity of $x_{cm}=0$ of Figure~\ref{fig:prob_cm_pos}g,h. The snapshots in Figure~\ref{fig:snapshots}g,h show stretching and alignment along the $x$-direction of the polymer-part between the fixed end point and the point where wrapping around the other fixed end appears. For such conformations, the major parts of the polymer were at $y<0$.
The~wrapping combined with the reversal of the polymer advective dynamics resulted in a slow lateral dynamics, resulting in a high probability with $y_{cm} < 0$. Yet, the average $\lla y_{cm} \rra$ was always positive, as shown in Figure~\ref{fig:ave_y_cm}, but decreased quickly with increasing $Wi$. A fit with the logarithmic Weissenberg-dependence $\lla y_{cm} \rra = - \beta r_0 \ln Wi$ yielded $\beta =4.9$ and $7.6$ for $L_p/L =1/2$ and $2$, respectively. Evidently, the drop is more pronounced for the stiffer polymer.

The~stronger conformational changes with increasing strength of the shear flow imply a shrinkage of the polymer. The~mean square radius of gyration shrinks by 10--15\%,  with respect to non-sheared~conformation.

\unskip

\subsection{Conformational Properties}
\vspace{-6pt}

\subsubsection{Normal Mode Expansion}

We studied the internal polymer conformational and dynamical properties via the mode amplitudes
${\bm A}_n(t)=(A_{nx}(t),A_{ny}(t))^T$ of the eigenfunction expansion of a polymer with fixed ends~\cite{wink:99},
\begin{equation}\label{modes}
{\bm r}_i = \frac{H(2i-N-1)}{2(N-1)}\hat {\bm x}
+ 2 \sum_{n=1}^{N-1}{\bm A}_n \sin(q_n[i-1]) \; ,
\end{equation}
with $\hat {\bm x}$ the unit vector along the $x$-axis and the wave numbers
$q_n=n \pi/(N-1)$ ($n=1,\ldots,N-1$). The~mode amplitudes are:
\begin{align}\label{comp}
{\bm A}_n  = \frac{1}{N-1}\sum_{i=1}^{N-1} {\bm S}_{i} \sin(q_n[i-1]),
\end{align}
in terms of the bead positions $S_{ix}=r_{ix}-(2i-N-1)H/2(N-1)$ and $S_{iy}=r_{iy}$ ($i=1,\ldots,N$).

\begin{figure}[H]
\centering
\includegraphics*[width=0.8\columnwidth,angle=0]{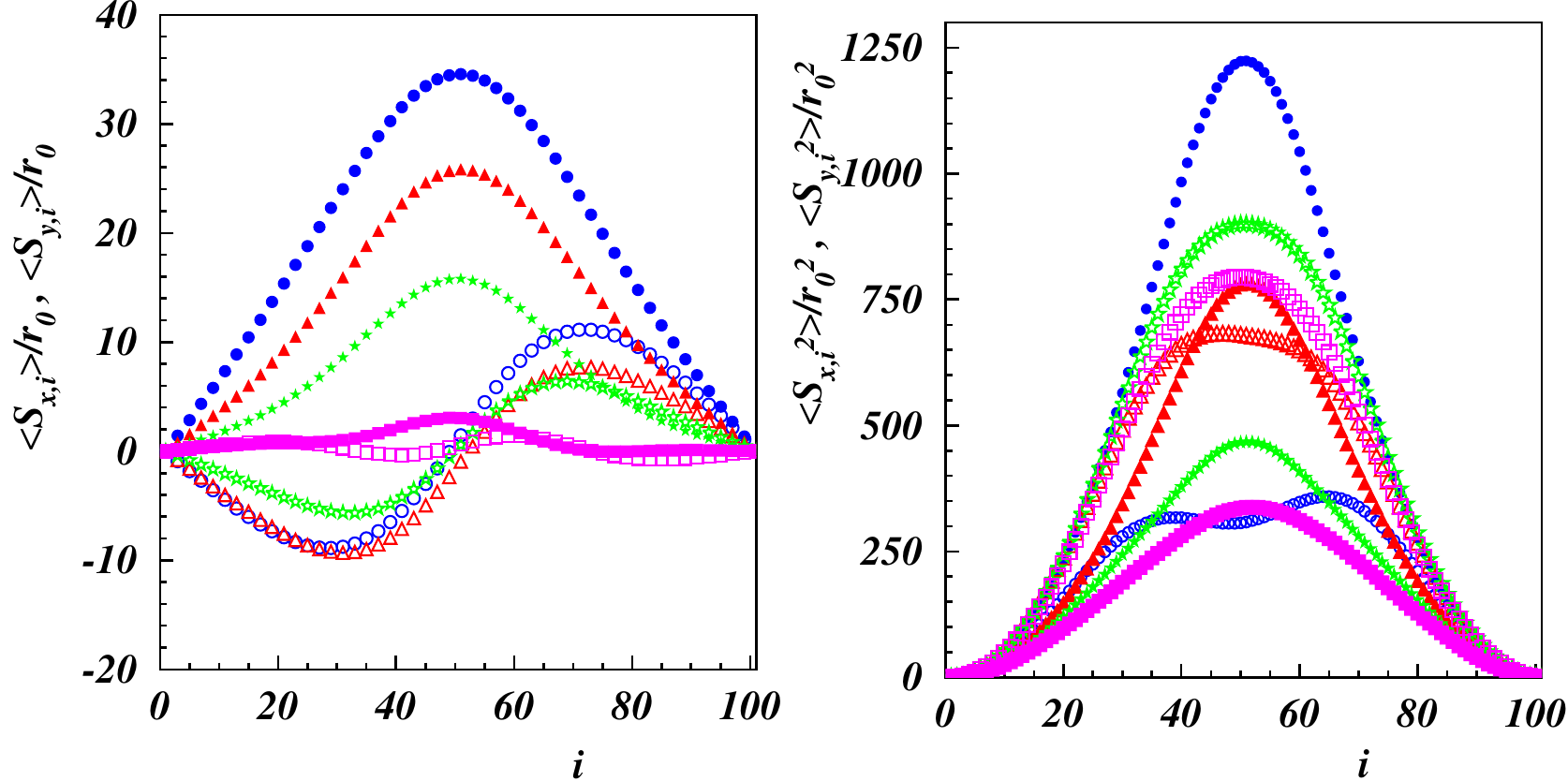}
\caption{Mean (\textbf{left}) and mean square (\textbf{right}) values of the components of the displacement $\bm S$ as a function of the bead index, $i$, of the semiflexible polymer with $L_p/L=2$ for $Wi=10~(\circ),$ $ 25~(\triangle),
50~(\star), 100~(\Box)$. Open symbols indicate $S_x$ and filled symbols $S_y$.
\label{fig:mean_s}
}
\end{figure}

The~mean, $\langle \bm S \rangle$, and mean square, $\langle \bm S^2 \rangle$, values of the components of $\bm S$ are displayed in Figure~\ref{fig:mean_s} for $L_p/L=2$. The~behavior was qualitatively similar for $L_p/L=1/2$. The~shape of $\langle S_x \rangle$ was centrosymmetric with respect the polymer center $i=50$. Any deviation was a consequence of statistical inaccuracy. The~magnitude $|\langle S_x \rangle|$ decreased significantly with the increasing Weissenberg number, and the mean was close to zero for $Wi =100$. The~component $\langle S_y \rangle$ was always positive and the largest amplitude was naturally in the polymer center. With increasing $Wi$, the amplitudes shrunk, and $\langle S_y \rangle$ was close to zero for all beads. The~mean square values $\langle S_y^2 \rangle$ (Figure~\ref{fig:mean_s}, right) decreased with increasing $Wi$. However, $\langle S_x^2 \rangle$ changed in a nonmonotonic manner, and the fluctuations were larger for $Wi=50$ than for $Wi=100$. This was a consequence of the symmetry-breaking shear flow with respect to the $x$-axis and the wrapping of the polymers around the tethering points.

\begin{figure}[H]
\centering
\includegraphics*[width=\columnwidth,angle=0]{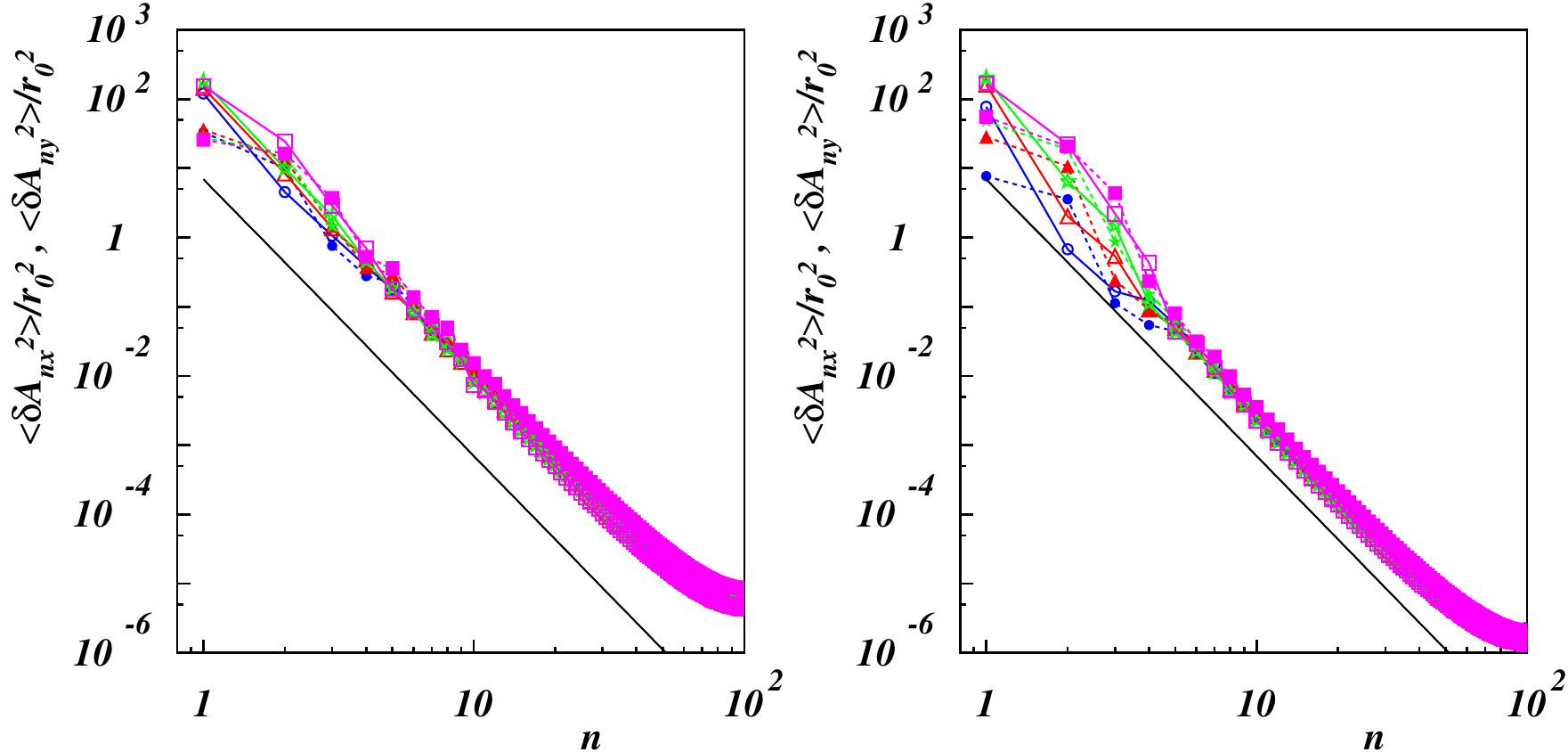}
\caption{Variance of the mode components
$A_{nx}$ (open symbols with solid lines)
and
$A_{ny}$ (solid symbols with dashed lines)
as functions of mode number $n$ for $L_p/L=0.5$ (\textbf{left}), $2$ (\textbf{right}) and
$Wi=10~(\bullet), 25~(\blacktriangle),
50~(\star), 100~(\blacksquare)$.
The~slope of the solid lines is $-4$.
\label{fig:mode_amp}
}
\end{figure}

Figure~\ref{fig:mode_amp} depicts the dependence of the variance of the mode amplitudes:
\begin{align}
\lla \delta \bm A_n^2 \rra = \lla \left( \bm A_n- \lla \bm A_n\rra \right)^2\rra,
\end{align}
on the mode number for the two stiffness values
and the various Weissenberg numbers. For small $Wi$, we obtained a $1/n^4$ dependence as characteristic for stiff polymers~\cite{harn:95,arag:85}. This even applied over a range of mode numbers $4< n \lesssim 40$ at higher shear. However, the variances of the lower-mode amplitudes deviated from this dependence, with a weaker mode-number dependence for small $n$, and~a rapid drop from mode $n=3$ to $n=4$. This reflected the large-scale conformational changes by the wrapping of the polymers.
\subsubsection{Dynamical Properties} \label{sec:dynamics}

\begin{figure}[H]
\centering
\includegraphics*[width=0.8\columnwidth,angle=0]{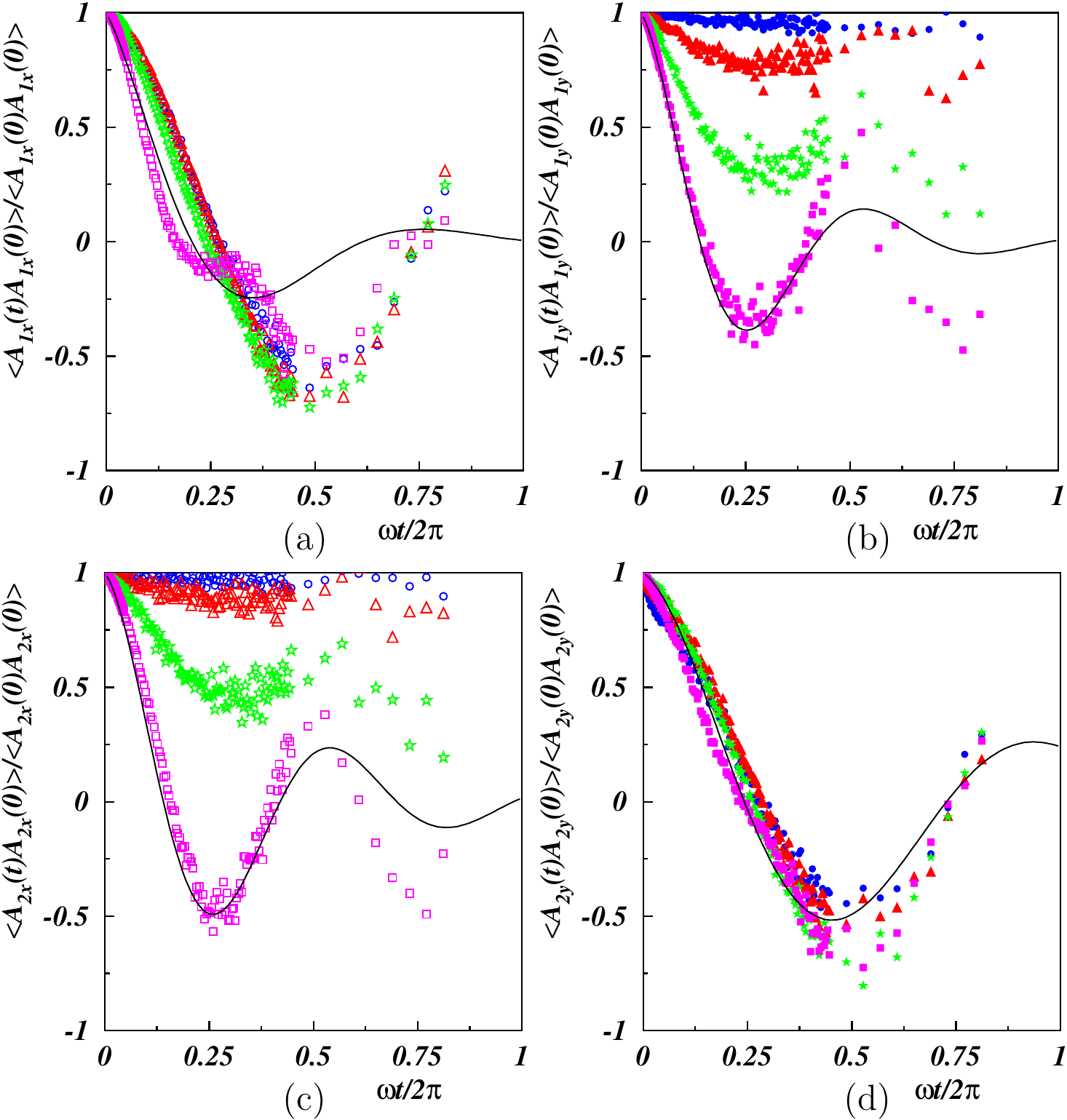}
\caption{ {Autocorrelation} function of the mode amplitudes for the modes $n=1$ (\textbf{top}) and $n=2$ (\textbf{bottom}) as a function of time along the $x$- (\textbf{left}) and $y$-direction (\textbf{right}). The~polymer stiffness is $L_p/L=2$ and the Weissenberg numbers
$Wi=10~(\circ), 25~(\triangle),
50~(\star), 100~(\Box)$. The~black solid lines are fits of a damped sinusoidal oscillation.
\label{fig:mode_auto_corr}
}
\end{figure}

Since the large-scale properties of the polymer were modified most by the shear flow, we~more closely considered the dynamics of the mode amplitudes for the modes $n=1$ and $2$ by the mode-autocorrelation function
$C_n(t) = \langle \bm A_n(t) \cdot \bm A_n(0) \rangle$.
Results for various Weissenberg numbers and the persistence length $L_p/L =2$ are presented in Figure~\ref{fig:mode_auto_corr}. The~simulation data were analyzed by fitting the exponentially damped periodic function:
\begin{align}
F(t) = e^{- \gamma t/T_0} \cos(\Omega \omega t),
\end{align}
where $T_0=2\pi/\omega$ is the period of the applied oscillation, $\gamma$ characterizes the damping, and $\Omega$ accounts for variations of $\omega$.
The~correlation $C_{1x}(t) =\langle A_{1x}(t)A_{1x}(0) \rangle$, in Figure~\ref{fig:mode_auto_corr}a, decreased at short times and increased again for $t/T_0 >1/2$. Fitting yielded a factor $\Omega \approx 1.2$, i.e., $C_{1x}(t)$ followed roughly the shear flow. At high Weissenberg numbers, an additional weak modulation appeared, which, however, did not change the primary frequency. The~decay of the correlation function $C_{1y}(t)=\langle A_{1y}(t)A_{1y}(0) \rangle$ showed a strong Weissenberg number dependence. As depicted in Figure~\ref{fig:mode_auto_corr}b,
$C_{1y}(t)$ depended only very weakly on time for $Wi=10$. With the increasing Weissenberg number, the decay rate, $\gamma$, increased, and the correlation function showed damped oscillations, which were most pronounced for $Wi=100$. Interestingly, the characteristic frequency was twice the externally applied frequency ($\Omega \approx 1.8$). The~correlation function $C_{2x}(t)$ exhibited a very similar time and Weissenberg number dependence as $C_{1y}(t)$ (cf. Figure~\ref{fig:mode_auto_corr}c), in terms of drop at the various $Wi$, as well as the characteristic frequency.
The~correlation $C_{2y}(t)$ (cf. Figure~\ref{fig:mode_auto_corr}d) was very similar to $C_{1x}(t)$, where the fitted expression closely followed the simulation data. Here, we found $\Omega \approx 1$, i.e., the external frequency, but the decay was stronger with $\gamma \approx 1.4$. Moreover, within both sets of correlation functions, the exponential decay depended only weakly on the Weissenberg number. It was easily recognized that there was an odd-even asymmetry between the $x$- and $y$-correlation functions. This asymmetry was also obtained for higher mode numbers.

Theoretical models of flexible and semiflexible polymers predict an exponential decay of the normal-mode correlation functions of the form $e^{-t/\tau_n}$, where the $\tau_n$ are the relaxation times~\cite{doi:86,wink:06}. In~the presence of external fields, e.g., shear flow, the time dependence is modified and the normal-mode correlation functions do not necessarily decay exponentially anymore~\cite{wink:10}, but the exponential factor is still determined by the relaxation times. We did find an initial exponential decay, but no clear mode-number dependence. Figure~\ref{fig:mode_amp} shows a deviation of the smaller mode numbers from the dependence $\tau_n \sim 1/n^4$, but the relaxation times of the modes $n=1$ and $2$ were still different. This is not reflected in Figure~\ref{fig:mode_auto_corr}, where the decay of $\langle A_{1x}(t)A_{1x}(0) \rangle$ and $\langle A_{2y}(t)A_{2y}(0) \rangle$ differs by  approximately a factor of $2.5$, which is roughly the ratio of the mode numbers. Hence, the oscillatory flow field strongly affected the relaxation times, at least at higher Weissenberg numbers. This reflects and is consistent with a rather nonlinear response of the polymer with respect to the external periodic~excitation.

\begin{figure}[H]

\centering
\includegraphics*[width=.97\columnwidth,angle=0]{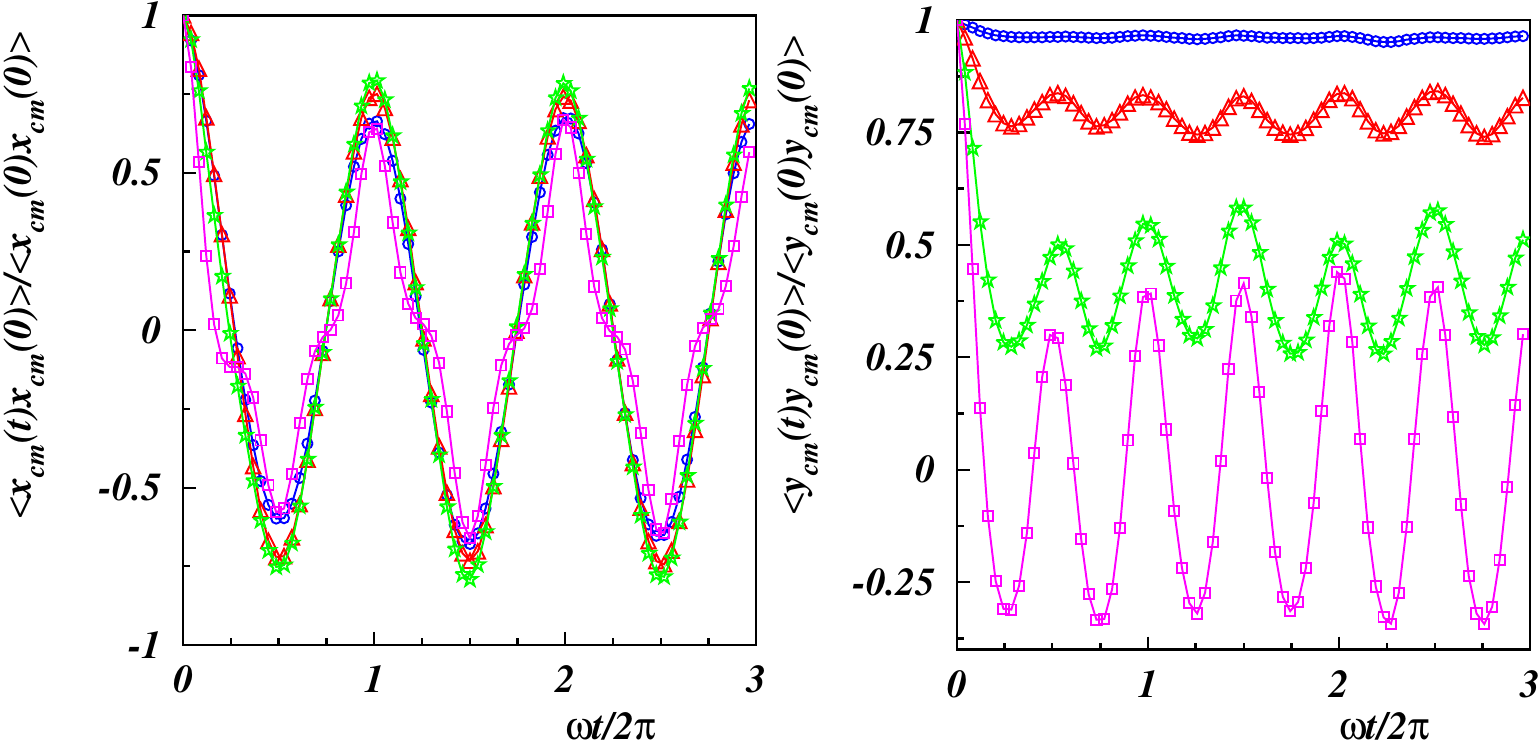}
\caption{Autocorrelation function of center-of-mass Cartesian coordinates
as a function of time along the $x$- (\textbf{left}) and $y$-direction (\textbf{right}). The~polymer stiffness is $L_p/L=2$ and the Weissenberg numbers
$Wi=10~(\circ), 25~(\triangle),
50~(\star), 100~(\Box)$.
\label{fig:auto_corr_cm}
}
\end{figure}

The~autocorrelation functions for the center-of-mass Cartesian coordinates displayed in Figure~\ref{fig:auto_corr_cm} help to understand the appearance of the frequency doubling. Evidently, both Cartesian components exhibited a periodic motion, with typically larger amplitudes for higher Weissenberg numbers. Clearly, the
$x$-component shows the same frequency as the applied flow, with some modulations
appearing for the highest value of $Wi$. In contrast, the $y$-component revealed frequency doubling for $Wi \gtrsim 25$. The~Supplementary Movies illustrate the different dynamical features of $x_{cm}$ and $y_{cm}$. In one period, the $x$-position of the center of mass, let us say, moved from the maximum positive value to the minimum value (half period) and, by the oscillation, back again to the maximum $x$-value. The~oscillation was similar to a complete ``rotation'' (period), and no difference between oscillation and rotation was visible. The~behavior of the amplitude of $y_{cm}$ was different. While in a cycle, $x_{cm}$ decreased, $y_{cm}$ increased first, reached its maximum for $x_{cm}=0$, and decreased then again. Instead of being able to complete a full cycle, the oscillation and the tethering constrained $y_{cm}$ to move back along a similar path as in the first half of the period, i.e., $\langle y_{cm}(t+T_0/2) \rangle$ increased, reached again a maximum and decreased then again to a value close to the initial value. Hence, $y_{cm}$ exhibited two maxima during a period, whereas $x_{cm}$ exhibited only one. As a consequence, the correlation function $\langle y_{cm}(t)y_{cm}(0) \rangle$
showed twice the frequency than the respective correlation function of the $x$-coordinate. The~primary reason was the self-avoidance by the tethering points, which forced a reversal of the dynamics along the $y$-direction.

\section{Summary and Conclusions}\label{sec:conclusions}

We have analyzed the nonequilibrium properties of semiflexible polymers confined in two dimensions with tethered ends exposed to oscillatory shear. The~applied Brownian multiparticle collision dynamics algorithm neglected hydrodynamic interactions. It coupled the polymer dynamics in a stochastic (Brownian) manner to the local flow field and allowed for fluctuations, since the average flow field was imposed only.

For small shear rates, low Weissenberg numbers ($Wi \lesssim 10$), the tethering of the ends led to a more-or-less linear oscillatory response, where the polymers moved back and forth like grass swaying in the wind. With increasing $Wi$, the polymers (partially) wrapped around the fixation points and a more complex, nonlinear response emerged. In fact, the wrapping significantly changed the polymer conformations, and the overall size, measured by the radius of gyration, shrunk. Dynamically, the~probability of the polymer center-of-mass position was largest on a spatial curve resembling a lima\c{c}on, although the distribution was inhomogeneous. Of course, this was a consequence of tethering and periodic excitation of the polymer, moving it cyclically back and forth, which led to an approximate rolling motion of the center-of-mass, the origin of the lima\c{c}on.

At high Weissenberg numbers, we found shear-induced modifications of the mode-spectrum of the mode-amplitude correlation functions. Since we considered semiflexible polymers only, the mode spectrum exhibited the characteristic $1/n^4$ dependence for low $Wi$. At high $Wi$, the lower modes, specifically the modes $n \leqslant 3$, showed a weaker $n$ dependence. This is reflected in a particular dynamical behavior, where a frequency doubling appeared for the normal-mode amplitudes along the $x$- and $y$-direction. However, the doubling appeared alternating for even and odd modes---the correlation function $\langle A_{1y}(t)A_{1y}(0) \rangle$ shows
twice the frequency of $\langle A_{1x}(t)A_{1x}(0) \rangle$, whereas $\langle A_{2x}(t)A_{2x}(0) \rangle$ is similar to $\langle A_{1y}(t)A_{1y}(0) \rangle$. The~frequency doubling is also reflected in the autocorrelation function of the $y$-coordinate of the center-of-mass position. It was a consequence of the hindered and truncated ``rotational'' motion of the polymer by the tethering points.

\RW{In the current simulations, hydrodynamic interactions (HI) have been neglected. Such interactions are of major importance for flexible polymers, since they significantly change the relaxation time spectrum, but are less relevant for semiflexible polymers, where they provide approximately logarithmic corrections only~\cite{harn:96}. Hence, we expected only minor differences between the presented results for free-draining polymers and simulations, accounting for HI as long as the polymer motion follows instantaneously the fluid flow.
}

Our studies revealed an intriguing behavior of tethered polymers under oscillatory flow, which affects their macroscopic rheological behavior. Specifically, the overall shrinkage of the polymer reduced the viscosity. The~consequences of the frequency doubling on macroscopic properties need to be further analyzed.
\vspace{6pt}

\authorcontributions{
R.G.W. and A.L. conceived the project. A.L. performed the simulations. R.G.W. and A.L. interpreted the data and wrote the paper. Both authors discussed the results and commented on the manuscript.}



\conflictsofinterest{
The~authors declare no conflict of interest.}


\bibliographystyle{mdpi}
\reftitle{References}

\end{document}